# Human Activity Recognition on Time Series Accelerometer Sensor Data using LSTM Recurrent Neural Networks

Chrisogonas O. Odhiambo
Computer Science & Engineering
University of South Carolina
Columbia, SC, USA
odhiambo@email.sc.edu

Sanjoy Saha
Pennington Biomedical Research
Center, Louisiana State University,
Baton Rouge, LA, USA
sanjoy.saha@pbrc.edu

Corby K. Martin
Pennington Biomedical Research
Center, Louisiana State University,
Baton Rouge, LA, USA
Corby.Martin@pbrc.edu

Homayoun Valafar
Computer Science & Engineering
University of South Carolina
Columbia, SC, USA
homayoun@cse.sc.edu

**Abstract** – *The use of sensors available through smart devices has pervaded everyday life in several applications including human activity monitoring, healthcare, and social networks. In this study, we focus on the use of smartwatch accelerometer sensors to recognize eating activity. More specifically, we collected sensor data from 10 participants while consuming pizza. Using this information, and other comparable data available for similar events such as smoking and medication-taking, and dissimilar activities of jogging, we developed a LSTM-ANN architecture that has demonstrated 90% success in identifying individual bites compared to a puff, medication-taking or jogging activities.*

**Keywords**: *Smartwatch, Accelerometer, Sensors, Artificial Intelligence, Machine Learning, LSTM, Human Activity Recognition, Eating, Bite, Food Intake*

## I.  INTRODUCTION

Accurately assessing health behaviors in humans is necessary to evaluate health risk and effectively intervene to facilitate behavior change, improve health, and reduce disease risk. Health behaviors, such as eating, smoking, exercise (e.g., jogging), and medication-taking are frequently assessed with subjective self-report methods, such as diaries, which participants complete throughout the day. The accuracy of self-report methods is poor, however, particularly for assessing food intake and physical activity [1]. Self-report methods are also burdensome for participants, particularly if health behaviors need to be assessed over the long term [2].

Mobile health technology, including sensors worn on the body, can be used to passively and remotely collect and transmit objective data. These objective data can be much more valid and reliable compared to self-report, particularly for exercises such as walking [3]. The passive collection and transmission of data to researchers or clinicians have other advantages, including a dramatic reduction in participant burden and the ability to process and provide feedback to participants automatically and in real-time or near real-time. This critical step provides a platform to develop and deliver ecological momentary interventions (EMI) [4] and just-in-time adaptive interventions (JITAI) [5]. EMI and JITAIs deliver intervention strategies that are customized to address the specific needs of individual participants as soon as these needs are detected. Participant needs are identified by evaluation of the objective data from the remote sensors in real-time or near real-time. Indeed, EMI and JITAI can provide more automated and cost-effective approaches to intervene and improve health behavior remotely while maintaining efficacy [6]–[8].

Here we report a novel application of Artificial Neural Networks to, objectively and automatically, identify and discriminate eating activity from three other activities namely smoking, medication-taking, and jogging using accelerometer data acquired from a smartwatch. Validation of the algorithm would make it possible to develop and deploy novel EMI and JITAI to improve these four health behaviors. Machine Learning algorithms have been used to achieve great results in developing practical solutions in multiple domains: health diagnosis [9]–[11], sports [12], [13], human activity recognition [14]–[16], among many others.

## II.  BACKGROUND AND METHOD

### A.  Previous and Related Work

Considering their rich array of sensors, the cost, accessibility, and ease of use, smartwatches have emerged as a compelling platform to study human activities unobtrusively. Smartwatches have been used as step-counters[17], sleep monitoring[18], diet monitoring[19] as well as general fitness tracking[20]. In the context of smoking, smartwatches have been demonstrated to be usable for in-situ study of smoking[21], [22] with high accuracy[21]–[24]. Smartwatches have been used to detect smoking gestures with 95% accuracy in a laboratory environment[25] and 90% in-situ detection of smoking[23]. The study of smoking has also been demonstrated to be more accurate when compared to self-report (90% versus 78%)[23], [26].

Wearable devices have been utilized in indirect observation of some activities with clear health implications, such as medication adherence in numerous ways, including (1) self-report the behavior via mobile devices[27], (2) sensors worn around the neck e.g. the SenseCam[28] was originally envisaged for use within the domain of Human Digital Memory to create a personal lifelog or visual recording of the wearer's life, which can be helpful as an aid to human memory, (3) multi-axis inertial sensors worn on wrists[29], [30], and (4) the use of commodity smartwatches [31]. In summary, all the approaches have collectively demonstrated the potential of smart sensors to promote the study of human activities but leave potential for improvement in performance, cost, convenience, and usability.

In this experiment we have utilized our collected data for eating activity and previously available data for smoking, medication-taking, and jogging recorded from human subjects. We have used this data to train and test an ANN capable of detecting eating behavior at the bite level.

### B. Data and the Acquisition Process

This study involved four sets of activities, namely eating, smoking, medication-taking, and jogging. The selection of these activities was influenced by several factors including the availability of data, and the degree of their similarity. In particular, smoking and medication-taking were included in this study to challenge the detection of eating by providing similar behaviors. Three of the four activities (eating, smoking, medication-taking) consist of hand-to-mouth, hand-off-mouth, and hand-on-mouth sequence of events and should, therefore, provide a reasonable assessment of the network's performance.

The following three behaviors were recorded using smartwatches (Polar M600, Asus Zenwatch, Motorola, TicWatch) running android operating system: eating pizza (without cutlery/utensils), medication-taking, and smoking. The participants wore the smartwatches on the wrist (right or left). Each watch is equipped with accelerometer sensor to measure the 3D acceleration, and gyroscope sensor to measure the 3D angular velocity. In this study, we only utilized the accelerometer sensor data sampled at a frequency of 25 Hz.

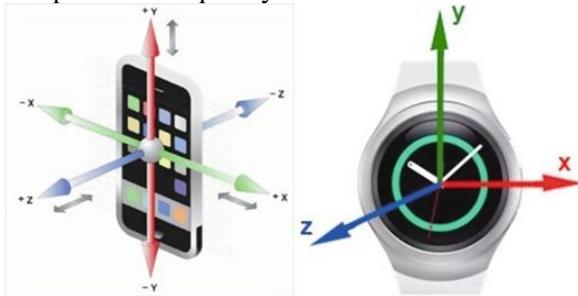

*Figure 1: An illustration of accelerometer axes on a typical smartphone and smartwatch*

The fourth behavior evaluated, jogging, relied on an open public data from Wireless Sensor Data Mining (WISDM) Lab (http://www.cis.fordham.edu/wisdm/) that was recorded using smartphone strapped to the waist location of the participant.

Data recording using the smartwatch was performed independently at the convenience of the participant. The paired phone was only necessary in data transmission. Both the phone and the watch were installed with Android App. On the watch side of the activity, our software performs the following: (1) initiate data collection, (2) allow the user to indicate start and end of an activity during data collection, (3) transmit data to the phone via Bluetooth. On the phone, the application performs the following: (1) receive data from watch, (2) uniquely name files and upload them to the research cloud repository. The cloud is a webservice that receives and logs data files from the phone.

### Data Pre-processing

Data pre-processing is a crucial step in the data-mining process. It involves data-filtering, replacement of the missing and outlier's values, as well as feature extraction/selection. The windowing technique is commonly used to extract features from raw data. The technique involves the segmentation of the sensor signals into small time blocks with overlapping [32], [33]. There are three types of windowing techniques namely: (i) sliding window, where the signals are divided into fixed-length blocks/windows, (ii) event-defined windows, where specific events are located/identified and restructured as successive data partitioning, and (iii) activity-defined windows, where partitioning is based on detection of specific activity changes [32]. This study applies the sliding window approach, which is well suited to real-time applications because it does not require any pre-processing.

The raw data logged at the cloud repository comprises zipped sensor files. Each zipped file comprises two files: actual raw data of tri-axial values and the corresponding timestamps. The second file contains annotation data provided by the user that identifies the approximate time of the activity of interest in the stream of raw data. These two files were processed using an in-house developed utility program that extracted gesture features from the raw data based on the approximate start/stop timestamps reported by the user in the second file. It is important to note that the timestamps reported by users are not directly useful for several reasons. First, a participant may report false start and stop times for a variety of reasons including simple human error. When accurately reported, the start and end portions of the signal will include hand movements that are unrelated to the activity of interest. For instance, hand movements related to clicking the start/stop buttons will also be included as part of the activity of interest. Therefore, to reduce noise in the data, increase its integrity and improve network model performance, the gestures were visually confirmed and trimmed at both 'tail' and 'head' ends of each activity by a "supervisor." Using this pre-processing pipeline, the usable output files were generated for training and testing of Machine Learning.

## C. Neural Network Platform and Architecture

Although in this study we evaluated numerous ML approaches, in the interest of brevity, here we report the most successful approach that consisted of LSTM-ANN. Long short-term memory (LSTM) is an artificial Recurrent Neural Network (RNN) architecture with feedback connections [34], [35]. An LSTM unit comprises a cell, an input gate, an output gate and a forget gate. The cell remembers values over arbitrary time intervals and the three gates regulate the flow of information into and out of the cell. LSTM networks find suitable applications in classifying, processing, and making predictions based on time series data. LSTM architectures address the vanishing gradient problem that can be encountered when training traditional RNNs. Figure 2 is an illustration of a typical LSTM cell where $x_t$ is the input vector to the LSTM unit, $h_t$ is the hidden state vector (or LSTM unit output vector), $c_t$ is the cell state vector, and $c_{t-1}$ is the cell input activation vector.

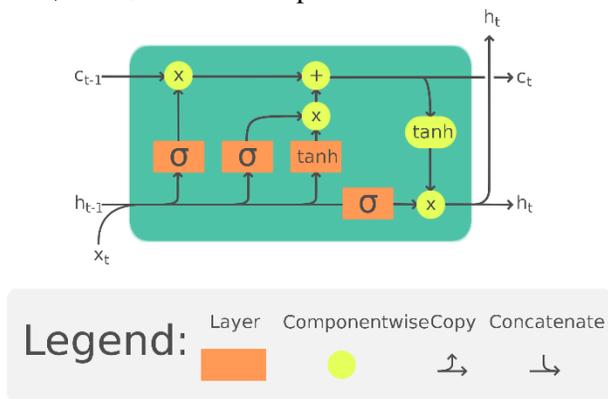

*Figure 2: The LSTM cell can process data sequentially and keep its hidden state through time. (By Guillaume Chevalier - File: The_LSTM_Cell.svg, CC BY-SA 4.0, https://commons.wikimedia.org/w/index.php?curid=109362147)*

In the classification of sequential data, it is common to ignore the sequential aspect of data and treat the data as if it were independently and identically distributed (*iid*), and subsequently apply a standard machine learning classification algorithm that is designed for *iid* data [36]. It is important to note that the temporal dependence in the sequence of data will determine the effectiveness of the approximation. This, however, depends on how data is pre-processed i.e., is the raw data passed to the classifier as-is, or are features computed from the time-series and then in turn passed to the classifier. The human activities of interest in this study – eating-pizza, medication-taking, smoking, and jogging – are each sequence of mini-activities whose temporal aspect adds important component in the overall activity recognition. For example, the eating activity is made up of a series of mini-actions namely pick-pizza slice, raise slice to mouth, bite-pizza, lower slice from mouth. The sub-activities, and their sequence, is important. The past sub-features are useful and relevant as far as recognition of the full activity is concerned. For this reason, LSTMs find most relevant application in recognition of this human activity time-series data.

Our implementation of LSTM architecture comprised 2 fully-connected and LSTM layers (stacked on each other) with 64 units each. Stacking LSTM hidden layers makes the model deeper, more accurate, making it suitable for complex activities. Although the output of the system could have consisted of a single neuron to denote the presence or absence of eating event, we decided to design a two-neuron output. One of the two neurons is designated to the presence of an eating event and the other neuron indicates the presence of the other activities (smoking, medication-taking, jogging, or others). This design was used in light of our future expansion of the system to detect an array of activities. The input of the network (window size) was experimentally determined to consist of 150 consecutive points representing six seconds of recording time. The window step was maintained at 10 datapoints, and hence an overlap computed as Window-size – Step-size; for example, in the case of window-size of 50 units, the overlap is given by (50 – 10 = 40). While we did not vary the step size, it is important to note that the smaller the step-size, the more real-time the data-series is. Smaller step-size improves performance, but increases window counts and slows down detection. The sliding window-with-overlaps process significantly transforms and reduces the training dataset. Further, the transformation assigns the most common activity (i.e., mode) as a label for the sequence; some windows comprise two or more activities, but the mode is considered the dominant or overriding activity. We transform the shape of our input into sequences of 150 rows, each containing x, y and z values (representing the accelerometer data). We also apply a one-hot encoding to the labels to transform them into numeric values that can be processed by the model [37]–[40].

## D. Training and Testing Procedure

As the first step in training of a network, we balanced the data by randomly repeating a continuous segment of the recorded data so that all classes have an approximately equal representation. The train/test datasets were generated from the balanced dataset by partitioning in 80:20 ratio, respectively. During the training process, the learning rate was set at 0.0025 and the model was trained for 50 epochs while keeping track of accuracy and error. The batch size was maintained at 1024. We applied L2 regularization (Ridge Regression) to the model. The L2 penalty/force removes a small percentage of weights at each iteration, ensuring that weights never become to zero. The penalty consequently reduces the chance of model overfitting.

**Evaluation**

The study used accuracy to evaluate classifiers performances. The metric measures the proportion of correctly classified examples. Accuracy can be expressed as follows:

$$Accuracy = \frac{TP+TN}{TP+TN+FP+FN} \quad (i)$$

Where TP (true positives) represent the correctly classified positive examples, TN (true negatives) represents the correctly classified negative examples, FP (false positives) represent negatives misclassified as positives, and FN (false negatives) represent positives misclassified as false. The accuracy measure does not take into account the bias arising from unbalanced datasets. Thus, the metric has a bias favor for the majority classes. For this reason, the study considered the following evaluation criteria: Precision, recall, F-measure, and specificity. Below are the formulae to compute the metrics:

$$precision = \frac{TP}{TP+FP} \quad (ii)$$

$$recall = \frac{TP}{TP+FN} \quad (iii)$$

F-measure is the combination of precision and recall. It is calculated as follows:

$$F\sim measure = \frac{(1+\beta^2).recall.precision}{\beta^2 recall+precision} \quad (iv)$$

where β is a weighting factor and a positive real number i.e. the weighted harmonic mean of precision and recall, reaching its optimal value at 1 and its worst value at 0. The beta parameter determines the weight of recall in the combined score. It is used to control the importance of recall/precision. To give more weight to the Precision, we pick a Beta value in the interval 0 < Beta < 1; To give more weight to the Recall, we pick a Beta Value in the interval 1 < Beta. We applied a score of 1.

Specificity is computed as follows:

$$specificity = \frac{TN}{TN+FP} \quad (v)$$

## III. RESULTS AND DISCUSSION

### A. Exploration and Visualization of the Data

Table 1 presents a summary of the total amount of data that was acquired or prepared for use in this study. In this table, the column denoted by Participants indicates the number of participants in each study, while Datapoints reports the total number of sampled data points collected across all participants. In essence, the number of Datapoints, when divided by 25 will correspond to the total duration of recording in seconds (e.g. a total of 3.03 hours of recording of the eating event). The column denoted as Patterns indicates the total number of presentations on the ANN that will contain the activity of interest (e.g. 5434 input patterns that contained an eating event). The large number of patterns is a result of a stream of data that presents different portions of the same activity (e.g., the same bite for the same participant) to ANN. Therefore, one singe gesture may consist of several input patterns to the ANN that consists of the activity of interest.

*Table 1: Summary of all the datasets used in the study*

| Activity | Datapoints | Patterns | Participants |
|---|---|---|---|
| Eating pizza | 272822 | 5434 | 10 |
| Jogging | 287461 | 5882 | 27 |
| Medication | 412798 | 3100 | 31 |
| Smoking | 62823 | 1279 | 15 |

As another critical step in annotation of data, it is important to visualize the data to become familiar with the intricate nature of each activity. This is a critical step for a "supervisor" to confirm and adjust the labeling of the outcome associated with each activity.

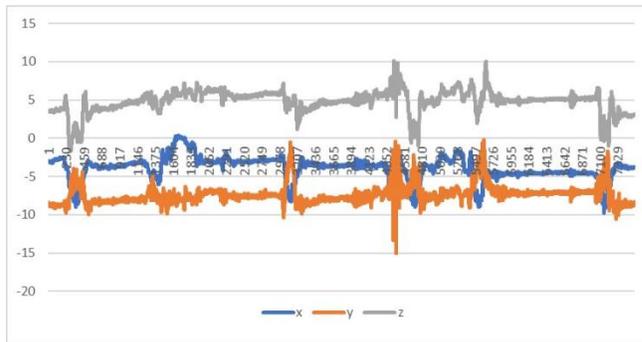

*Figure 3: An entire recording of consuming pizza. This recording consists of 6 bites that in total took 147 seconds.*

Figure 4, Figure 5, Figure 6 and Figure 7 are visualizations of the tri-axial data for individual eating, smoking, medication-taking, and jogging activities, respectively. It is important to note that eating, smoking, and jogging consist of repetitious sub-activities (a bite, a puff, a step) that are displayed in each of the corresponding graphs. Medication-taking activity, on the other hand, is composed of several sub-activities that appear in some temporal sequence and collectively appear only once. For instance, one medication-taking activity may consist of the sequence of opening a pill bottle, dispensing a pill, taking the pill, drinking water, putting everything back (pill bottle and water bottle, etc.). Previous work provides a detailed dissection of the medication-activity with an annotation of the sub-activities [14].

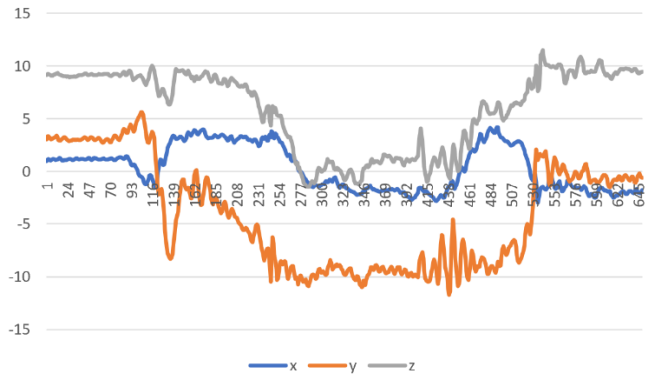

Figure 4: An eating gesture consisting of a single bite of pizza.

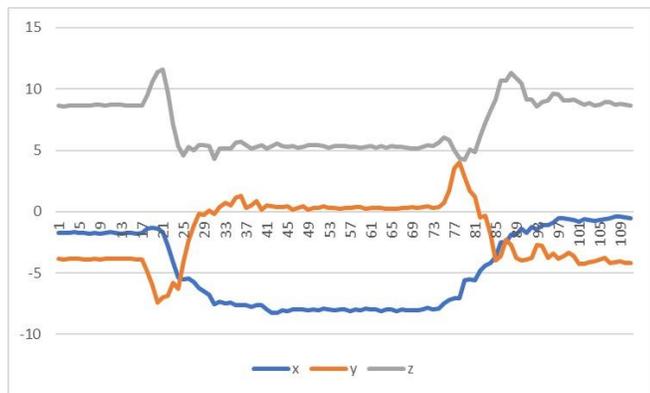

Figure 5: Single smoking gesture

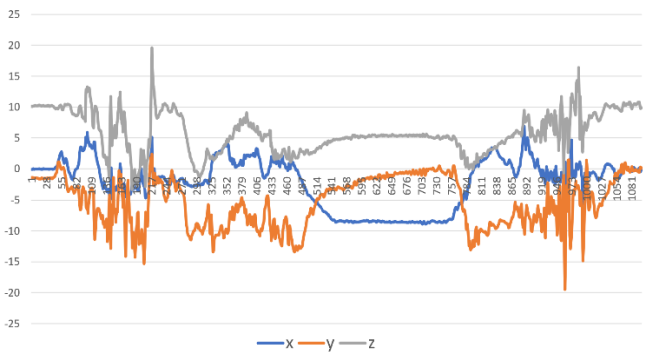

Figure 6: A single medication-taking gesture that consists of multiple sub-events (opening bottle, dispensing pill, taking pill, drinking water, etc.)

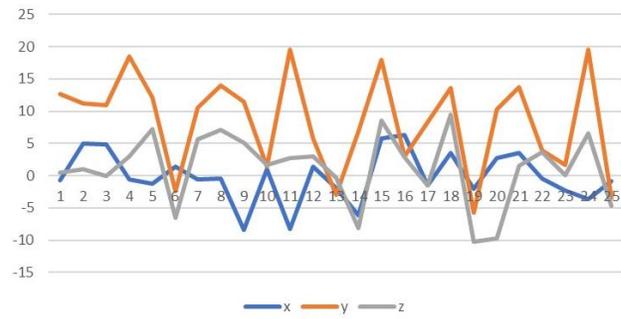

Figure 7: Three consecutive steps during jogging activity.

### B. Training of ANN

The training of the LSTM-ANN proceeded upon the creation of the labeled input files (with balancing) using the hyper-parameters described in sections II.C and II.D. In this study we utilized TensorFlow/Keras as the primary ANN simulation platform. The training and testing process took place on Google Colab(oratory). The Colab is a Google Research cloud service with a web IDE for python as well as computing services. The specifications of the computing resources we used were as follows: Intel(R) Xeon(R) CPU @ 2.20GHz, 13GB RAM, and 40GB of storage. The storage is upgradable to 108GB. The training of the system with 50 epochs required approximately 3 hours of dedicated compute time. The training and testing loss functions are illustrated in Figure 8. Based on the results shown in this figure, it is clear that network has successfully learned the presented classification task.

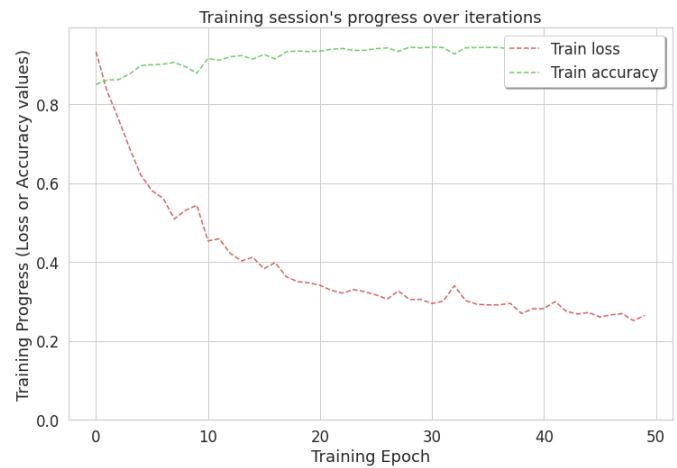

Figure 8: Training plot for the window size of 150 units. The configuration resulted the best performance among the different models.

### C. Testing Results

Following the successful training of the LSTM-ANN, we embarked on testing of the system using 20% of the randomly selected original data set. Table 2 lists the results for the test set using different metrics. Based on the results shown in this table, the system achieved a performance of approximately 90% success. To explore the full nature of the misclassifications, the confusion table (shown in Figure 9) was examined. The accuracy, precision, recall, F-measure, and specificity, as described by the formulae *i, ii, iii, iv and v*, respectively, are presented the Table 2.

*Table 2: Summary metrics for the best performing configuration of window size 150*

| Activity | Precision | Recall | F-Measure | Specificity | Accuracy |
|---|---|---|---|---|---|
| Eating | 0.89 | 0.97 | 0.93 | 0.96 | 0.96 |
| Other | 0.94 | 0.92 | 0.93 | 0.99 | 0.98 |

| True label | Eating | 5286 | 288 |
|---|---|---|---|
| | Other | 792 | 13587 |
| | | Eating | Other |
| | **Predicted label** | | |

*Figure 9: Confusion matrix for the window size of 150 units. The configuration resulted the best performance among the different models.*

## IV. CONCLUSION

Automated detection of eating activity is of critical importance in relation to obesity and healthy weight management. Automatic detection of eating sessions will help to remove the burden of self-reporting from the participants and therefore, provide a simpler way of tracking eating events. In this report, we demonstrated successful identification of individual bites with an accuracy of approximately 90% when tested against activities that significantly resemble to eating. In particular, smoking and medication-taking will share the common mini-gestures of hand-to-mouth, hand-on-mouth, and hand-off-mouth. Although not presented here, our initial investigation has confirmed very little confusion between eating and jogging. Therefore, we speculate the accuracy of the system to increase notably if other natural daily activities are included in our training and testing sets due to their dissimilarity to the eating gesture.

Although in this work we have achieved a reasonably high detection of the eating gesture, a number of additional investigations can be initiated to increase the performance and usability of the system. First, as an ultimate objective, we aim to develop one application that can decipher numerous human activities to establish correlative or causative relationship between activities. For instance, eating at 1:00 PM may lead to a cigarette smoking soon after. The ability to monitor the temporal relationship between these two events would be very useful. To accomplish this, we need to engage in a formal investigation of the optimal viewable window size to and ANN that will be sufficient to successfully decipher between all activities of interest. Furthermore, there exists some inherent parallel between human activities and the principles of written language. To fully leverage this parallel analogy, human activities need to be examined in the more fundamental fashion by identifying the mini gestures that are the basis set of all complex activities. Here we will resort to some of the previous work [15] in order to understand the mini-gesture decomposition of the eating activity in relation to other similar activities such as smoking.

## REFERENCE

[1] N. V. Dhurandhar *et al.*, "Energy balance measurement: when something is not better than nothing," *Int. J. Obes. (Lond).*, vol. 39, no. 7, pp. 1109–1113, Jul. 2015.

[2] C. Höchsmann and C. K. Martin, "Review of the validity and feasibility of image-assisted methods for dietary assessment," *Int. J. Obes. (Lond).*, vol. 44, no. 12, pp. 2358–2371, Dec. 2020.

[3] K. Kastelic, M. Dobnik, S. Löfler, C. Hofer, and N. Šarabon, "Validity, Reliability and Sensitivity to Change of Three Consumer-Grade Activity Trackers in Controlled and Free-Living Conditions among Older Adults," *Sensors 2021, Vol. 21, Page 6245*, vol. 21, no. 18, p. 6245, Sep. 2021.

[4] K. E. Heron and J. M. Smyth, "Ecological momentary interventions: incorporating mobile technology into psychosocial and health behaviour treatments," *Br. J. Health Psychol.*, vol. 15, no. Pt 1, pp. 1–39, Feb. 2010.

[5] I. Nahum-Shani *et al.*, "Just-in-Time Adaptive Interventions (JITAIs) in Mobile Health: Key Components and Design Principles for Ongoing Health Behavior Support," *Ann. Behav. Med.*, vol. 52, no. 6, pp. 446–462, May 2018.

[6] C. K. Martin, A. C. Miller, D. M. Thomas, C. M. Champagne, H. Han, and T. Church, "Efficacy of SmartLoss, a smartphone-based weight loss intervention: results from a randomized controlled trial," *Obesity (Silver Spring).*, vol. 23, no. 5, pp. 935–942, May 2015.

[7] C. K. Martin, L. A. Gilmore, J. W. Apolzan, C. A. Myers, D. M. Thomas, and L. M. Redman, "Smartloss: A Personalized Mobile Health Intervention for Weight Management and Health Promotion," *JMIR Mhealth Uhealth 2016;4(1)e18 https//mhealth.jmir.org/2016/1/e18*, vol. 4, no. 1, p. e5027, Mar. 2016.

[8] L. M. Redman *et al.*, "Effectiveness of SmartMoms, a Novel eHealth Intervention for Management of Gestational Weight Gain: Randomized Controlled Pilot Trial," *JMIR mHealth uHealth*, vol. 5, no. 9, Sep. 2017.

[9] B. E. Odigwe, J. S. Eyitayo, C. I. Odigwe, and H. Valafar, "Modelling of Sickle Cell Anemia Patients Response to Hydroxyurea using Artificial Neural Networks," Nov. 2019.

[10] L. Zhao, B. Odigwe, S. Lessner, D. Clair, F. Mussa, and H. Valafar, "Automated analysis of femoral artery calcification using machine learning techniques," *Proc. - 6th Annu. Conf. Comput. Sci. Comput. Intell. CSCI 2019*, pp. 584–589, Dec. 2019.

[11] B. E. Odigwe, F. G. Spinale, and H. Valafar, "Application of Machine Learning in Early Recommendation of Cardiac Resynchronization Therapy," Sep. 2021.

[12] A. A. Aguileta, R. F. Brena, O. Mayora, E. Molino-Minero-re, and L. A. Trejo, "Multi-sensor fusion for activity recognition—a survey," *Sensors (Switzerland)*, vol. 19, no. 17. MDPI AG, 01-Sep-2019.

[13] V.-Y. H, H. P, S. J, V. T, and S. H, "Reliable recognition of lying, sitting, and standing with a hip-worn accelerometer," *Scand. J. Med. Sci. Sports*, vol. 28, no. 3, pp. 1092–1102, Mar. 2018.

[14] C. Odhiambo, P. Wright, C. Corbett, and H. Valafar, "MedSensor: Medication Adherence Monitoring Using Neural Networks on Smartwatch Accelerometer Sensor Data."

[15] C. O. Odhiambo, C. A. Cole, A. Torkjazi, and H. Valafar, "State transition modeling of the smoking behavior using lstm recurrent neural networks," *Proc. - 6th Annu. Conf. Comput. Sci. Comput. Intell. CSCI 2019*, pp. 898–904, Dec. 2019.

[16] W. Jiang and Z. Yin, "Human activity recognition using wearable sensors by deep convolutional neural networks," in *MM 2015 - Proceedings of the 2015 ACM Multimedia Conference*, 2015, pp. 1307–1310.

[17] V. Genovese, A. Mannini, and A. M. Sabatini, "A Smartwatch Step Counter for Slow and Intermittent Ambulation," *IEEE Access*, vol. 5, pp. 13028–13037, 2017.

[18] X. Sun, L. Qiu, Y. Wu, Y. Tang, and G. Cao, "SleepMonitor," *Proc. ACM Interactive, Mobile, Wearable Ubiquitous Technol.*, vol. 1, no. 3, pp. 1–22, Sep. 2017.

[19] S. Sen, V. Subbaraju, A. Misra, R. K. Balan, and Y. Lee, "The case for smartwatch-based diet monitoring," in *2015 IEEE International Conference on Pervasive Computing and Communication Workshops, PerCom Workshops 2015*, 2015, pp. 585–590.

[20] A. Pfannenstiel and B. S. Chaparro, "An investigation of the usability and desirability of health and fitness-tracking devices," in *Communications in Computer and Information Science*, 2015, vol. 529, pp. 473–477.

[21] N. Saleheen *et al.*, "PuffMarker: A multi-sensor approach for pinpointing the timing of first lapse in smoking cessation," in *UbiComp 2015 - Proceedings of the 2015 ACM International Joint Conference on Pervasive and Ubiquitous Computing*, 2015, pp. 999–1010.

[22] A. L. Skinner, C. J. Stone, H. Doughty, and M. R. Munafò, "StopWatch:The preliminary evaluation of a smartwatch-based system for passive detection of cigarette smoking," *Nicotine and Tobacco Research*, vol. 21, no. 2. Oxford University Press, pp. 257–261, 01-Feb-2019.

[23] C. A. Cole, D. Anshari, V. Lambert, J. F. Thrasher, and H. Valafar, "Detecting Smoking Events Using Accelerometer Data Collected Via Smartwatch Technology: Validation Study.," *JMIR mHealth uHealth*, vol. 5, no. 12, p. e189, Dec. 2017.

[24] C. A. Cole, D. Anshari, V. Lambert, J. F. Thrasher, and H. Valafar, "Detecting Smoking Events Using Accelerometer Data Collected Via Smartwatch Technology: Validation Study," *JMIR mHealth uHealth*, vol. 5, no. 12, p. e189, Dec. 2017.

[25] "(16) (PDF) Recognition of Smoking Gesture Using Smart Watch Technology." [Online]. Available: https://www.researchgate.net/publication/315720056_Recognition_of_Smoking_Gesture_Using_Smart_Watch_Technology. [Accessed: 08-Nov-2019].

[26] L. E. Wagenknecht, G. L. Burke, L. L. Perkins, N. J. Haley, and G. D. Friedman, "Misclassification of smoking status in the CARDIA study: A comparison of self-report with serum cotinine levels," *Am. J. Public Health*, vol. 82, no. 1, pp. 33–36, 1992.

[27] S. L. West, D. A. Savitz, G. Koch, B. L. Strom, H. A. Guess, and A. Hartzema, "Recall Accuracy for Prescription Medications: Self-report Compared with Database Information," *Am. J. Epidemiol.*, vol. 142, no. 10, pp. 1103–1112, Nov. 1995.

[28] D. Byrne, A. R. Doherty, G. J. F. Jones, A. F. Smeaton, S. Kumpulainen, and K. Järvelin, "The SenseCam as a Tool for Task Observation," 2008.

[29] M. Straczkiewicz, N. W. Glynn, and J. Harezlak, "On Placement, Location and Orientation of Wrist-Worn Tri-Axial Accelerometers during Free-Living Measurements," *Sensors (Basel).*, vol. 19, no. 9, May 2019.

[30] A. Doherty *et al.*, "Large scale population assessment of physical activity using wrist worn accelerometers: The UK biobank study," *PLoS One*, vol. 12, no. 2, Feb. 2017.

[31] I. Maglogiannis, G. Spyroglou, C. Panagopoulos, M. Mazonaki, and P. Tsanakas, "Mobile reminder system for furthering patient adherence utilizing commodity smartwatch and Android devices," in *Proceedings of the 2014 4th International Conference on Wireless Mobile Communication and Healthcare - "Transforming Healthcare Through Innovations in Mobile and Wireless Technologies", MOBIHEALTH 2014*, 2015, pp. 124–127.

[32] F. Attal, S. Mohammed, M. Dedabrishvili, F. Chamroukhi, L. Oukhellou, and Y. Amirat, "Physical Human Activity Recognition Using Wearable Sensors," *Sensors 2015, Vol. 15, Pages 31314-31338*, vol. 15, no. 12, pp. 31314–31338, Dec. 2015.

[33] P. Casale, O. Pujol, and P. Radeva, "Human Activity Recognition from Accelerometer Data Using a Wearable Device," *Lect. Notes Comput. Sci. (including Subser. Lect. Notes Artif. Intell. Lect. Notes Bioinformatics)*, vol. 6669 LNCS, pp. 289–296, 2011.

[34] Y. Yu, X. Si, C. Hu, and J. Zhang, "A Review of Recurrent Neural Networks: LSTM Cells and Network Architectures," *Neural Comput.*, vol. 31, no. 7, pp. 1235–1270, Jul. 2019.

[35] S. Hochreiter and J. Schmidhuber, "Long short-term memory," *Neural Comput.*, vol. 9, no. 8, pp. 1735–1780, Nov. 1997.

[36] N. Twomey *et al.*, "A Comprehensive Study of Activity Recognition Using Accelerometers," 2018.

[37] "TensorFlow-on-Android-for-Human-Activity-Recognition-with-LSTMs/human_activity_recognition.ipynb at master · curiously/TensorFlow-on-Android-for-Human-Activity-Recognition-with-LSTMs." [Online]. Available: https://github.com/curiously/TensorFlow-on-Android-for-Human-Activity-Recognition-with-LSTMs/blob/master/human_activity_recognition.ipynb. [Accessed: 09-Nov-2021].

[38] "Human Activity Recognition using LSTMs on Android | TensorFlow for Hackers (Part VI) | Curiously - Hacker's Guide to Machine Learning." [Online]. Available: https://curiously.com/posts/human-activity-recognition-using-lstms-on-android/. [Accessed: 09-Nov-2021].

[39] Y. Zhao, R. Yang, G. Chevalier, X. Xu, and Z. Zhang, "Deep Residual Bidir-LSTM for Human Activity Recognition Using Wearable Sensors," *Math. Probl. Eng.*, vol. 2018, 2018.

[40] "guillaume-chevalier/HAR-stacked-residual-bidir-LSTMs: Using deep stacked residual bidirectional LSTM cells (RNN) with TensorFlow, we do Human Activity Recognition (HAR). Classifying the type of movement amongst 6 categories or 18 categories on 2 different datasets." [Online]. Available: https://github.com/guillaume-chevalier/HAR-stacked-residual-bidir-LSTMs. [Accessed: 09-Nov-2021].